\documentclass[12pt,aps,a4paper,nofootinbib]{revtex4}
\usepackage{amsmath}
\usepackage{graphicx}
\usepackage[usenames,dvipsnames]{color}

\newcommand{\re}{\ref}
\newcommand{\be}{\begin{equation}}
\newcommand{\ee}{\end{equation}}

\newcommand{\la}{\label}
\newcommand{\ber}{\begin{eqnarray}}
\newcommand{\eer}{\end{eqnarray}}

\newcommand{\bs}[1]{\ensuremath{\boldsymbol{#1}}}

\begin{document}

\title{Calculating reactions with use of no--core shell--model states\footnote{Phys. Rev. C {\bf 99}, 034620 (2019),
DOI: 10.1103/PhysRevC.99.034620}}

\author{  Victor D. Efros
  }

\affiliation{
National Research Centre "Kurchatov Institute", 123182 Moscow, Russia\\
and National Research Nuclear University MEPhI, 115409 Moscow, Russia 
}
 
\date{\today}
 
\begin{abstract}
A scheme to compute reactions is described that uses the Slater determinants constructed of oscillator 
orbitals. Simple linear equations are suggested for this purpose and shown to be efficient in model examples.   
A universal
 method to evaluate the required matrix elements is given.
\end{abstract}

\bigskip

\maketitle
\section{introduction}

Last decades a big progress has been made in the ab initio description of $p$--shell nuclei in the framework
of the no--core shell--model (ncsm) method~\cite{Bar}.
As already was pointed out~\cite{Nav}  a challenging task is to
extend this method to describe reactions. 

Work in this direction was done in Refs.~\cite{qn,bar,nq,qnrh,pap,s}.
In Ref.~\cite{qn} the resonating--group method has been employed.  
The bound--state wave functions of the heavier of reaction partners were taken in the form
of the 
ncsm expansions over the Slater determinants.  The  bound--state wave functions of the lighter  of
reaction partners were treated as an expansion over oscillator functions 
in the Jacobi coordinates. The coupled
set of integro--differential equations describing
relative motion of fragments in a reaction has been obtained.  Kernels of the equations were derived
in the fashion of Ref.~\cite{Nav} 
for the cases when the mass numbers of the 
lighter reaction partner are equal to one~\cite{qn}, two~\cite{nq}, and three~\cite{qnrh}.

Ncsm   pseudo--states of fragments were subsequently included in the calculations. This 
resulted in increase of
the number of the
 integro--differential equations  while convergence with respect to adding 
pseudo--states
proved to be too slow in certain cases. In this connection, it was suggested~\cite{bar} to supplement
the resonating group ansatz with  a set of states 
belonging to the ncsm pseudo--spectrum of the whole system. To realize this,
the $R$--matrix approach was used and the Bloch--Schr\"odinger equation
was solved in conjunction with the resonating group description. The required additional matrix elements (ME) 
were obtained in a way similar to that of Ref.~\cite{Nav}. In the latter paper, at calculating the spectroscopic function
of a nucleus
the heavier cluster and the lighter one
were treated 
in the same manner as mentioned above
and formulae for the ME were obtained
separately for the cases of lighter clusters consisting of one, two, three, and four nucleons. In this sequence, the formulae 
become increasingly complicated, and the same refers to the above mentioned resonating group ME. 
The convergence issue 
was scrutinized in those investigations and stability of the reaction observables, at least at the qualitative level, was established.

Other shell--model approaches to describe reactions such as the no--core Gamow shell model, e.g., ~\cite{pap}, and
the $J$--matrix one~\cite{s} are also being developed.

Our purpose is to propose a simple and universal extension of ncsm to calculate reactions.  The case of 
two--fragment reactions is considered. 
 In the next section, simple linear
equations suitable for this purpose are described. Unlike the resonating--group 
approach, they do not involve antisymmetrization between 
nucleons belonging to different reaction partners. They prove to lead to rather precise results in model examples. 

In Secs. 3--5 
the
issue of calculating the ME entering the equations is addressed. In the difference to the above mentioned approach
\cite{qn,bar,nq,qnrh}
we employ the ncsm--type Slater determinant expansions for both fragments and we do not use Jacobi coordinates. We provide
simple formulae to calculate the required ME. They are universal, i.e., the same for fragments 
consisting of different numbers of nucleons.

Below the notaion like $\Psi_{nlm}({\bf X})$ refers to the eigenfunctions of the oscillator Hamiltonian
$-(1/2)\Delta_{\bf X}+(1/2)X^2$. These eigenfunctions are assumed to be 
normalized to unity, $n$ denotes the radial quantum number, and
$l$ and $m$ denote the angular momentum quantum numbers. In the nucleon orbital case, ${\bf X}={\bf r}/r_0$ 
where ${\bf r}$ is the  nucleon position vector and $r_0$ denotes the nucleon oscillator radius.

.
\section{Scheme for computing reactions}

\subsection{Formulation}

Continuum wave functions we shall deal with are the following. 
Assume that only two--fragment reaction channels are open. Below
quantites referring  to such a reaction channel, say~$i$, will be supplied with the corresponding
subscript.  
Denote the 
wave number  and the orbital momentum of relative motion of fragments  as $k_i$  and
$l_i$. Denote the mass numbers of fragments pertaining to a reaction channel as $A_{1i}$ and $A_{2i}$ and  the vector 
connecting their centers of masses prior to inter--fragment antisymmetrization   as ${\bs\rho}_i$,
\be {\bs\rho}_i=(A_{1i})^{-1}\sum_{k=1}^{A_{1i}}{\bf r}_k-(A_{2i})^{-1}\sum_{k=A_{1i}+1}^{A_{1i}+A_{2i}}{\bf r}_k
\la{rho}\ee
where ${\bf r}_k$ are nucleon positions. 
The following radial functions of relative motion of fragments will be employed,
\ber f^{(0)}_i(\rho_i)=\left(\frac{A_{1i}A_{2i}}{k_i}\right)^{1/2}\,
\frac{{\tilde G}_{l_i}(k_i,\rho_i)}{\rho_i},
\qquad
f^{(1)}_i(\rho_i)=\left(\frac{A_{1i}A_{2i}}{k_i}\right)^{1/2}\,
\frac{F_{l_i}(k_i,\rho_i)}{\rho_i}
\la{z1}\eer
where
$F_l$ is the regular Coulomb function and ${\tilde G}_l$ is obtained from the irregular Coulomb function~$G_l$ 
by means of a regularization at small
distances. One may set, for example, \mbox{${\tilde G}_l(k,\rho)=g_l(\rho)G_l(k,\rho)$} with 
\be g_l(\rho)=[1-\exp(-\rho/\rho_{cut})]^{2l+1}\la{reg}\ee 
where $\rho_{cut}$ is a parameter.

Define the "surface functions"~$\varphi_i$,
\be \varphi_i=
\Bigl[\left[\phi_1^{I_1}(1,\ldots,A_{1i})\phi_2^{I_2}(A_{1i}+1,\ldots,A_{1i}+A_{2i})\right]_{S}
Y_{l}({\hat{\bs \rho}}_i)\Bigr]_{JM}.\la{sf}\ee
Here $\phi_1^{I_1M_1}$ and $\phi_2^{I_2M_2}$ are bound--state wave functions of fragments. 
They contain, respectively,
 nucleons with
the numbers from $1$ to $A_{1i}$ and from $A_{1i}+1$ to $A_{1i}+A_{2i}=A$ where  $A$ is the total number 
of nucleons in a system. These are internal wave functions  
depending on Jacobi vectors and  possessing given total momenta and their projections. The latter quantities are
denoted as $I_1,M_1$ and $I_2,M_2$. These wave functions possess also given  parities and isospins and 
they are the eigenfunctions of corresponding internal Hamiltonians.
The brackets~$[\ldots]$
represent couplings to the spin~$S$  of the two fragments and  to
the total spin~$J$ and its projection~$M$.

We shall deal with the channel functions of two types
 denoted as $\psi_{i}^{(0)}$ and $\psi_{i}^{(1)}$. They are of the  form
\be \psi_{i}^{(0),(1)}={\cal A}_i\varphi_if^{(0),(1)}_i(\rho_i)\Psi_{000}({\bf{\bar R}}_{cm})\la{ch}\ee
where the functions~$f^{(0),(1)}_i$ are defined in Eq.~(\re{z1}). 
Here ${\cal A}_i$ is the inter--fragment antisymmetrizer,
\be  {\cal A}_i=\nu_i^{-1/2}\sum_{P}(-1)^{\pi(P)}P,\la{a}\ee
where $\nu_i$ is the number of channels in the configuration space which are associated with a given reaction channel~$i$.
These channels in the configuration space 
correspond to different distributions of nucleons over the fragments so that $\nu_i=A!/(A_{1i}!A_{2i}!\upsilon)$
where $\upsilon=2$ if the fragments are identical, and $\upsilon=1$ otherwise. 
The number of terms in the sum is $\nu_i$ and   the permutations~$P$ are such that with their help one 
obtains all the  channels  in the configuration
space from one of them. The quantity~$\pi(P)$ is either zero or one depending on the parity of a 
permutaition.
As mentioned above, in Eq.~(\re{ch}) $\Psi_{000}$ is the 
ground--state harmonic oscillator
function and \mbox{${\bf{\bar R}}_{cm}={\bf R}_{cm}\sqrt{A}/r_0$ where
${\bf R}_{cm}$} is the center--of--mass vector of the whole system. 

Define approximate (or trial, or truncated) continuum wave functions 
\be \Psi_j=\psi_{j}^{(0)}+\sum_{i=1}^na_{i}^j\psi_{i}^{(1)}+\sum_{k=1}^{N-n}b_{k}^j\chi_k,\la{an}\ee
$1\le j\le n$. The functions~$\psi_{m}^{(0),(1)}$ are defined in Eq.~(\re{ch}) and they correspond
to open reaction channels whose number is denoted as $n$.
The functions
$\chi_k$ are short--ranged. They  are linear combinations of the Slater determinants constructed
of the oscillator orbitals. Their choice is discussed below. 
%


The expansion coefficients~$a_{i}^j$ and $b_k^j$  are to be found.
One has
 $a_i^j=-K_{ij}$ where $K_{ij}$ is the $K$--matrix.
 The $S$--matrix sought for is 
\be S=(iK+I)(iK-I)^{-1}.\la{sk}\ee 
For brevity we
rewrite the ansatz (\re{an}) as 
\be \Psi_j=\zeta_0^j+\sum_{i=1}^Nc_{i}\zeta_i,\qquad 1\le j\le n\la{Psi}\ee
where $\zeta_0^j=\psi_{j}^{(0)}$, and at $1\le i\le n$ one has $\zeta_i=\psi_{i}^{(1)}$.  
At $i=n+k$ one has $\zeta_i=\chi_k$. 
To find the coefficients of the expansions of the type of Eq.~(\re{an}) 
simple equations of the form
\be \sum_{i=1}^NA_{ki}c_i=B^{(j)}_k,\qquad k=1,\ldots,N\la{leq}\ee
will be employed.

In the case of  small systems,
the Hulth\'en--Kohn variational method  is traditionally applied.  
It leads to the equations of Eqs.~(\re{leq}) form
with $A_{ki}=(\zeta_k,[H-E]\zeta_i)$ and $B^{(j)}_k=-(\zeta_k,[H-E]\zeta_0^j)$.
Here and below $H$ is an internal Hamiltonian and $E$ is the energy 
of the whole system. 
The disadvantage of such equations in our case is that they thus include the "free--free" ME like
$\left(\psi_k^{(1)},[H-E]\psi_l^{(0),(1)}\right)$  which represent a class of ME additional 
to bound--bound and bound--free ME and which are more involved than the latter ones.

In view of this, in Ref.~\cite{ze} (subsec. 5 of Sec. 1 there) another set of equations has been suggested.
Formulating them, let us  take into account that 
the quantities like $[H-E]\psi_{i}^{(0),(1)}$ are localized. This is seen when one interchanges  
$H-E$  with the ${\cal A}_i$ operator entering Eq.~(\re{ch}) and then writes as usual 
\be H-E=(H_1-\epsilon_1)+(H_2-\epsilon_2)+[T_{rel}+{\bar V}_{ext}^{coul}(\rho_i)-
E_{rel}]+[V_{ext}^{nucl}+V_{ext}^{coul}
-{\bar V}_{ext}^{coul}(\rho_i)]\la{dec}\ee
where $H_1$ and $H_2$ are internal Hamiltonians of the fragments, $\epsilon_1$~and~$\epsilon_2$
are fragment eigenenergies, $T_{rel}$ is the  operator of kinetic energy of the fragment  relative motion,  
$E_{rel}$ is the energy of this relative motion,
${\bar V}_{ext}^{coul}=Z_1Z_1e^2/\rho_i$ is a subsidiary 
potential that reproduces the large--distance Coulomb  inter--fragment interaction, 
and $V_{ext}^{nucl}$ and  $V_{ext}^{coul}$ are nuclear and Coulomb
inter--fragment interactions. 

The third term in Eq.~(\re{dec}) acts on the relative motion functions
$f^{(0),(1)}_i(\rho_i)Y_{lm}({\hat{\bs \rho}}_i)$ and the corresponding contribution
 decreases exponentially as $\rho_i$ increases. 
The contribution of the nuclear interaction $V_{ext}^{nucl}$ entering  the last term  also decreases 
exponentially as $\rho_i$ increases while the difference $V_{ext}^{coul}-{\bar V}_{ext}^{coul}$
of the Coulomb interactions there decreases as $\rho_i^{-2}$.

Since in the class of  trial $\Psi_j$ of Eq.~(\re{Psi}) form the state $[H-E]\Psi_j$ is thus localized,  to select
the "best" of the $\Psi_j$ it is natural to require~\cite{ze} that $[H-E]\Psi_j$ is orthogonal to~$N$ lowest 
localized states~$\chi_k$,
that are of the type of the $N-n$  states~$\chi_k$ entering Eq.~(\re{an}).  I.e., we set 
$\left(\chi_k,[H-E]\Psi_j\right)=0$ with $1\le k\le N$. These are the  equations having the form of Eq.~(\re{leq}) with 
\be A_{ki}=\left(\chi_k, [H-E]\zeta_i\right),\qquad B^{(j)}_k=-\left(\chi_k, [H-E]\zeta_0^j\right).\la{mo}\ee
These equations include bound--bound and bound--free ME only. As pointed out 
in~\cite{ze} one more advantage of such type equations is that in ME entering them
antisymmetrization with respect to nucleons
belonging to different fragments may be omitted. These equations were not studied numerically
in~\cite{ze}  and, to our knowledge,    
they were not employed in the literature
in practical scattering calculations. The corresponding comments in~\cite{ze} were general ones and were not specially
intended for applications of the type of the present paper. 
However, these equations are very suitable for our present purpose of extending  ncsm to describe reactions.
In the subsequent sections the calculation of bound--free ME entering them 
is addressed. 

In an independent paper~\cite{lad} the least--square type equations have been suggested to solve the problem.
It was pointed out there that the equations of the form (\re{leq}) and (\re{mo}) are the limiting case of the least--square method.
In our case, when the matrix of the $H-E$~operator in the Slater determinant basis is large
and sparce, this general least--square method seems to be less efficient
 than these equations.

The above trial wave functions include regularization parameters like $\rho_{cut}$ in Eq.~(\re{reg}). Let us discuss
their choice. One may suggest that in the case of a sufficiently accurate calculation there should exist ranges of these
parameters such that reaction amplitudes are nearly independent of their choice within these ranges. If so, the optimal
values of these parameters are those which belong to these ranges, see the examples below. This prescription
corresponds to the fact that the
convergent values of reaction amplitudes are independent of such parameters. 

Let us discuss the choice of the short--range basis states~$\chi_k$ entering Eqs.~(\re{an}), (\re{Psi}), and (\re{mo}).
In our case, they are many--body oscillator states. 
In the case of projection equations of Eqs.~(\re{mo}) type, the 
results of a calculation are completely determined by a linear space the $\chi_k$~states 
span. If the problem is considered in the space of all the corresponding oscillator states with the numbers
of many--body oscillator quanta up to some value then it may occur
that the convergence is not reached unless the space is very large.
But the existing experience on solving large systems of linear equations,
see, e.g., \cite{saad}, suggests that it may be sufficient to solve the problem  only in a modest size Krylov subspace 
of that large space.
This subspace is spanned by the states $\phi,PH\phi,\ldots,(PH)^{N_0}\phi$ where $P$ is the projector onto the
mentioned space of oscillator states and $\phi$ is a pivot state belonging to this space.
 The $N_0$ value is expected ~\cite{saad} not to exceed several hundreds to reach convergence.

The set of  $(PH)^n\phi$ states is "ill--posed". An equivalent
 basis set in the above Krylov subspace which is convenient for performing calculations is the 
Lanczos basis set that starts  from the $\phi$~state. I.e., the $\chi_k$~states  are the corresponding Lanczos
states at such a choice. 

In our case, it is convenient to have  the pivot $\phi$~state in the form of a superposition of the Slater determinants. 
When $N_0$ is sufficiently high, the results of such type calculations~\cite{saad}
are often not sensitive to the choice of the $\phi$~state.  In  the literature in
a number of cases   this state is chosen to be an approximate solution, if
known in advance, 
of the 
set of 
linear equations to be solved.  In other cases it is chosen to be the right--hand side of linear
equations to be solved. 

The latter choice may be realized 
approximately in our problem as follows. One first constructs directly a complete subset, 
with not too high maximal number of many--body 
oscillator quanta, of basis oscillator functions
having the center of mass of the whole system in the lowest oscillator state and having given total angular momentum 
quantum numbers. Then expanding the right--hand side term 
$(H-E)\psi_{j}^{(0)}\equiv(H-E)\zeta_0^j$ over this subset with the help of the
method described below one gets a $\phi$~state sought for. 

The choice of  an approximate solution 
of our problem as the $\phi$~state may also be realized at use of the 
mentioned subset of basis oscillator functions. Then the whole problem is to be solved directly with the corresponding 
limited number
of these basis functions. The linear combination  $\sum_{k=1}^{N-n}b_{k}^j\chi_k$ from Eq.~(\re{an})
obtained in
the framework of this approximation
 may then be employed as the $\phi$~state in subsequent more extensive calculations. (At constructing such a pivot 
state it is natural to take  
all the corresponding states with the total number of oscillator quanta up to
some value as the $N-n$ states~$\chi_k$ entering Eq.~(\re{an}). Then 
the question arises how to choose $n$ extra  $\chi_k$~states on which the projecting in Eqs.~(\re{mo}) is
to be done. See the model example below in this connection.) 
  
It is desirable to deal with the $\chi_k$~states for which the center of mass of the whole system is
in the lowest oscillator state and the total angular momentum  and its projection have given values.
Most calculations of the ME below refer to this case. The $M$--scheme approach is adopted below
so that the projection of the total angular momentum  is  given anyway. As to the two other mentioned properties, 
the discussed pivot $\phi$~state has these properties and hence the same is valid for all the other
$\chi_k$~states if the exact arithmetic is assumed.

However, in certain cases 
these properties along with the nice properties of the Lanczos basis may be destroyed because of round--off errors
 even in computations with the quadrupole precision. Despite this,  the $\chi_k$~states 
provided by the above procedures still may be employed as basis states.
Of course, this is to be done 
without relying on the mentioned properties.  The angular momentum and the center--of--mass quantum numbers
then will be recovered in the total wave function once convergence of a calculation is reached. 
In this version,  ME more general than those mentioned above are required.
We want to note that  
the evaluation of 
these ME is discussed in short at the end of Sec. 5 while
the rate of convergence in this regime is to be investigated.

Another way to diminish the influence of round--off errors discussed in the literature, e.g., \cite{shir}, is the following. 
At calculating
spectra,   the  $\beta[H_{cm}-(3/2)\hbar\omega]$ operator, $H_{cm}$ being the oscillator center-of--mass Hamiltonian, 
with a large $\beta$ constant is added to the internal Hamiltonian.
This shifts above the spurious center--of--mass
 excitations \cite{gl}. If one imagines the influence of round--off errors as an action of
a perturbation added to the Hamiltonian then it may be concluded that the added operator, indeed, diminishes coupling 
due to round--off errors between center--of--mass excited and unexcited states.
If this picture is valid then it may be reasonable to 
diagonalize the Hamiltonian in the Lanczos basis, i.e.,  to get a number of the
standard ncsm solutions for the whole $A$--nucleon system, and to
use these solutions as $\chi_k$, cf. \cite{bar}. The corresponding Lanczos, or Krylov, subspace is, however, different from such 
subspaces discussed above and, unlike those subspases, it is not related to the inhomogenous equations to be solved.

In fact, in general no reasons are known for 
convergence to be faster in the case of the Hulth\'en--Kohn type equations  than in the case of Eqs.~(\re{leq}) 
with the coefficients (\re{mo}).
Below a good convergence is demonstrated in the case of the latter equations in model examples.

Reaction parameters obtained from various dynamics equations are sometimes
considered to be the first approximation and
presumably improved  values of these parameters are then obtained taking the
stationary values of the corresponding
Hulth\'en--Kohn functionals to be final results. 
However, in reality these functionals do not exist in the literal sense in the case when
bound--state wave functions of reaction participants are not exact.
The derivations  of  the Hulth\'en--Kohn variational principle for this case we know in the 
literature~\cite{gw,del} are inconclusive. 
In view of this, only in cases when the accuracy of a calculated reaction wave function is considerably lower than the accuracy  
of fragment wave functions entering it 
one could hope that the values of reaction parameters thus obtained are more accurate than the
original ones.  Besides, the  Hulth\'en--Kohn functionals include undesirable free--free ME.  For these reasons, we refrain from 
this improvement procedure.

The calculation can also be performed in the incoming and outgoing wave representation. In such a case 
the following radial functions of the relative motion of fragments are employed instead of functions (\re{z1}),
\be f^{\pm}_i(\rho_i)=\left(A_{1i}A_{2i}/k_i\right)^{1/2}\,
{\tilde H}_{l_i}^{(\pm)}(k_i,\rho_i)/\rho_i\la{f}\ee
where 
\[  {\tilde H}_{l}^{(\pm)}(k,\rho)={\tilde G}_l(k,\rho)\pm iF_l(k,\rho).\]
Correspondingly, the following channel functions are used instead of functions (\re{ch}),
\be \psi_{i}^{\pm}={\cal A}_i\varphi_if^{\pm}_i(\rho_i)\Psi_{000}({\bf{\bar R}}_{cm}).\la{ch1}\ee
The representation of the form 
\be \Psi_j=\psi_{j}^-+\sum_{i=1}^na_{i}^j\psi_{i}^++\sum_{k=1}^{N-n}b_k^j\chi_k,\la{an1}\ee
$1\le j\le n$,
similar to Eq.~(\re{an}) is used to obtain the
approximate continuum wave functions. The $\chi_k$~terms are the same as in Eq.~(\re{an}).
One has
 $a_i^j=-S_{ij}$ where $S_{ij}$ is the $S$--matrix.
Let us rewrite the ansatz of Eq.~(\re{an1}) in the form of  Eq.~(\re{Psi}) where now
$\zeta_0^j=\psi_{j}^{-}$, and at $1\le i\le n$ one has $\zeta_i=\psi_{i}^+$.  As above, $\zeta_i=\chi_k$ at $i=n+k$. 
With this notation, the equations of the same form (\re{leq}) and (\re{mo}) 
are applicable in the present case. 


In Secs. IV and V bound--free ME entering Eqs.~(\re{mo}) are calculated. Let us use the notation
\be Z_{I_1I_2SJM}^{nl}=\Bigl[\left[\phi_1^{I_1}(1,\ldots,A_{1})\phi_2^{I_2}(A_{1}+1,\ldots,A_{1}+A_{2})\right]_{S}
\Psi_{nl}({\bs{\bar \rho}})\Bigr]_{JM}\Psi_{000}({\bf{\bar R}}_{cm}).\la{y}\ee
These quantities are obtained from those of Eqs.~(\re{sf}) and (\re{ch}), or (\re{sf})  and~(\re{ch1}), via
disregarding antisymmetrization and replacing the function of relative motion of fragments with an
oscillator function. 
The quantity ${\bs{\bar \rho}}$ here
equals  \mbox{$[A_1A_2/(A_1+A_2)]^{1/2}{\bs \rho}/r_0$} where $\bs \rho$ is 
the inter--fragment distance of the type of Eq.~(\re{rho}). 
ME we deal with below are of the form 
\be \left(\chi_k,{\hat O}Z_{I_1I_2SJM}^{nl}\right)\la{mey}\ee
where $\chi_k$ are as above and ${\hat O}$ is a scalar  operator.

\subsection{Examples}

To verify the convergence of the method of calculating scattering at use of Eqs.~(\re{mo})
consider first
the model problem of the $s$--wave scattering of a particle by the potential~$-V_0\exp(-r/R)$, $V_0>0$. It
was employed in the literature~\cite{lad,mo} to study the Hulth\'en--Kohn and least--square methods. 
The problem allows an analytic solution, see, e.g.,~\cite{Ga}.

The precise values of the phase shift~$\delta$ can be obtained as follows. Let us use the notation~$k$ 
for the wave number and denote $\hbar^2/(\mu R^2)$, $\mu$ being the particle mass, as~$E_0$. 
 Define
\be F=\sum_{n=0}^\infty a_n,\qquad a_n=\frac{-2V_0/E_0}{n(n+2ikR)}a_{n-1}\la{ex}\ee
with $a_0=1$. Then $\tan\delta={\rm Im}F/{\rm Re}F$.

We want to find out whether Eqs.~(\re{leq}) and (\re{mo}) lead to the exact 
solution of the problem.
Let us take, for example, $V_0=E_0$ and $kR=0.1$. The number of bound states in the system is determined solely 
by a $V_0/E_0$ value. At the above condition, there exists one bound state and
its binding energy~$E_b$ is small compared to~$V_0$. One has
$E_b/E_0\simeq 0.013$. Once $V_0/E_0$ and $kR$ values are given 
the phase shift~$\delta$
is independent
of~$R$ and thus it refers to a family of potentials. Let $R$ be, say, 1.5 f.
If, in addition, $\mu$ is chosen to be
the reduced mass of the two--nucleon system then at the chosen $V_0/E_0$ and $kR$ values the
potential has the depth~$V_0$ about~37~MeV and  the center--of--mass scattering energy is about 0.2~MeV.   
At the chosen $V_0/E_0$ and $kR$ values
the value of $\tan\delta$ \mbox{equals -0.9798735} that is exact
in all the listed digits

In the present case the expansion (\re{an}) reads as
\be \psi(r)=\frac{\sin kr}{kr}+\tan\delta\left[(1-e^{-r/R_{cut}})\frac{\cos kr}{kr}\right]+
\sum_{m=1}^{N-1}b_m\chi_m(r)\la{mex}\ee  
where $R_{cut}$ is a parameter similar to $\rho_{cut}$ in ({\re{reg}). 
We choose the localized functions $\chi_m(r)$ to be the following,
\be \chi_m(r)=d^{-3/2}[m(m+1)]^{-1/2}L_{m-1}^2(r/d)e^{-r/(2d)}\la{basl}\ee
where $L_n^2(x)$ are the Laguerre polynomials. The basis set is orthonormalized. It is equivalent 
to the  $r^{m-1}\exp[-r/(2d)]$~set but provides a higher stability at solving the equations.
The ME required in the problem are calculated analytically in Appendix A. 

The calculation involves
two non--linear parameters, $R_{cut}/R$ and $d/R$. The $R_{cut}/R$ quantity is taken 
to be 1.0 in all the cases and the
results are insensitive to its choice within a wide range around this value. Eqs.~(\re{leq}) and (\re{mo}) we solve correspond
 to projecting
the Schr\"odinger equation onto the first $N$ basis functions from the set (\re{basl}). 

In Table 1 the values of $\tan\delta$ obtained are listed  for various numbers~$N-1$ of the 
basis functions~(\re{basl}) retained in Eq.~(\re{mex}). The results are shown for the choice $d/R=0.4$
and for a less favorabale choice $d/R=1.0$. It is seen that in the former case 
rather accurate results emerge already at small $N$~values. 
\begin{table}[h]
\caption{Dependence of $\tan\delta$ on the number of the short--range functions retained in Eq.~(\re{mex})}
\vspace{0.2cm}
\begin{tabular}{|c|c|}
\hline
\multicolumn{2}{|c|}{$d/R=0.4$}\\
\hline
$N-1$&$\tan\delta$\\
\hline
2&-0.9803861\\
\hline
4&-0.9810342\\
\hline
6&-0.9807419\\
\hline
8&-0.9800367\\
\hline
10&-0.9798684\\
\hline
20&-0.9798734\\
\hline
22&-0.9798735\\
\hline
\end{tabular}
\hspace{2cm}
\begin{tabular}{|c|c|}
\hline
\multicolumn{2}{|c|}{$d/R=1.0$}\\
\hline
$N-1$&$\tan\delta$\\
\hline
2&-0.7998181\\
\hline
8&-0.9783943\\
\hline
20&-0.9802085\\
\hline
30&-0.9798741\\
\hline
48&-0.9798735\\
\hline
\end{tabular}
\end{table}
The convergence patterns are rather similar in the whole energy region of interest.

 The following feature has been observed in our calculations.  
In Eqs.~(\re{leq}) and (\re{mo}) the last of the functions (\re{basl}) corresponded to $m=m_{last}=N$.
In place of  it,  now let us  use  
  functions $\chi_{m_{last}}$ 
of Eq.~(\re{basl}) form  
with $m_{last}=N+1$, or $N+2$, etc. 
It occurs that the results thus obtained 
are quite insensitive to the choice of the $m_{last}$~value. This is illustrated in Table 2 at the $d/R=0.4$ choice.

\begin{table}[h]
\caption{Dependence of $\tan\delta$ on the choice of $\chi_{m_{last}}$, see the text}
\vspace{0.2cm}
\begin{tabular}{|c|c|}
\hline
\multicolumn{2}{|c|}{$N=11$}\\
\hline
$m_{last}$&$\tan\delta$\\
\hline
11&-0.9798684\\
\hline
12&-0.9798479\\ 
\hline    
13&-0.9798180\\
\hline     
14&-0.9797937\\ 
\hline   
15&-0.9797773\\
\hline     
16&-0.9797675\\
\hline   
17&-0.9797620\\
\hline
\end{tabular}
\hspace{2cm}
\begin{tabular}{|c|c|}
\hline
\multicolumn{2}{|c|}{$N=21$}\\
\hline
$m_{last}$&$\tan\delta$\\
\hline
21&-0.9798734\\
\hline
22&-0.9798734\\ 
\hline    
23&-0.9798734\\
\hline   
24&-0.9798734\\
\hline    
25&-0.9798734\\ 
\hline   
26&-0.9798735\\
\hline     
27&-0.9798735\\     
\hline
\end{tabular}
\end{table}

This means that, at a sufficiently large number of basis functions retained in Eq.~(\re{mex}),  
the emerging scattering phase is in general
insensitive to the space onto which the Schr\"odinger equation is projected. At the same time, the projecting onto higher basis
states would be helpful if unphysical zero eigenvalues of the $A_{ki}$ matrix (\re{mo}) occur 
in an energy region
of interest.

The choice  of both the above potential and the basis (\re{mex}) aimed to verify with a high confidence
convergence of the method in general.
Now let us clarify features of the corresponding calculations 
in the case when the oscillator
basis is used and the number of retained basis functions is moderate. For this purpose, the model of Ref. \cite{su},
see also \cite{es},
is convenient. The model originates from a hyperspherical description of a three--particle system in the lowest hypermomentum
approximation. Effectively, it reduces to 
photodisintegration of a bound state of a nucleon with  the half--integer orbital momentum 3/2 in a  well $V(r)$. 

The well is a Gauss one,
$V(r)=-V_0\exp[-(r/R_0)^2]$ with 
$V_0=75$~MeV and $R_0=2.5$~f. The binding energy is about 3.5~MeV.

We seek for the final state continuum wave function in the form
\be  
\psi_E(r)=
\frac{J_{l+1/2}(kr)}{(kr)^{1/2}}-\tan\delta\left[1-\exp\left[-(r/R_{cut})^2\right]\right]^{l+1/2}
\frac{N_{l+1/2}(kr)}{(kr)^{1/2}}+\sum_{n=1}^{N}c_nR_{nl}(r)\la{csp}
\ee
where $l=5/2$, $R_{cut}$ is a parameter, and $R_{nl}$ are the radial oscillator functions. All the terms here
behave as $r^l$ at $r\rightarrow0$. The oscillator radius has been chosen to 
roughly optimize the calculation of the inital state binding energy
and it equaled~2~f. We retain seven lowest oscillator functions in the calculation, $N=7$ in Eq.~(\re{csp}).
Equations of the same form as in the preceding example are used to determine the $\tan\delta$ and  $c_n$ coefficients.
Exact results to compare with may be obtained, in particular,
within the same approach at use of sufficiently large $N$ values in Eq.~(\re{csp}).

\begin{figure}[ht]
\centerline{\includegraphics*[scale=0.5]{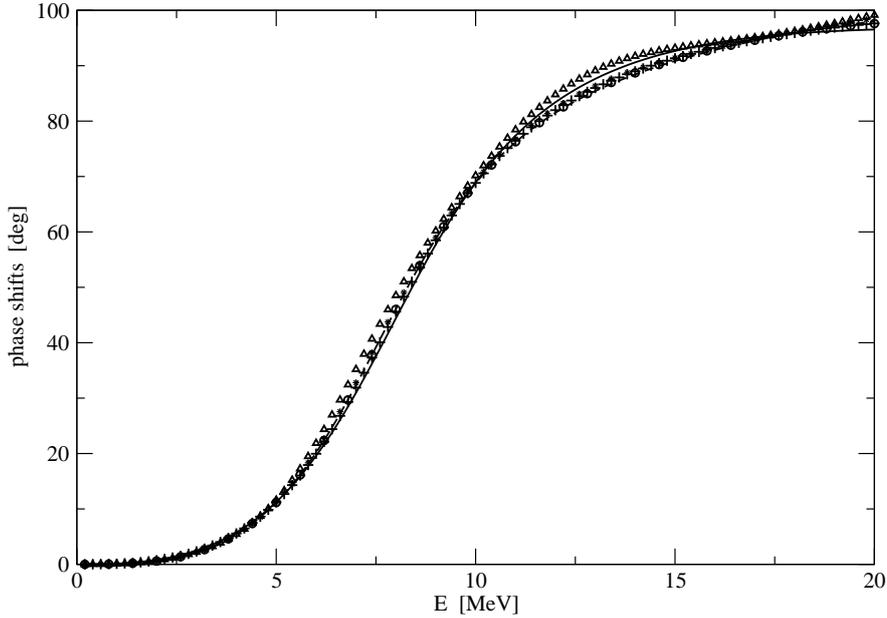}}
\caption{Comparison of the exact phase shift (full line) with those calculated with use of the oscillator basis. 
Pluses, circles, dotted line, stars, and triangles correspond to Eq.~(\re{csp}) with $N=7$ and with
$R_{cut}=2.5$~f, 3~f, 4~f, 5~f, and 6~f respectively.} 
\end{figure}

In Fig.~1 the phase shifts obtained are plotted along with the exact phase shift. The calculations were done 
with the values of the $R_{cut}$ parameter in Eq.~(\re{csp}) of 2.5~f, 3~f, 4~f, 5~f, and 6~f. It is seen that the results
pertaining to all the $R_{cut}$ values  except for 6~f are nearly indistinguishable and are close to the exact phase shift. 
This agrees with the above 
guess  that stability of calculated
reaction observables with respect to such type parameters indicates a sufficient accuracy of a calculation.
The results at $R_{cut}=6$~f are a little less accurate. At $R_{cut}=2$~f the phase shift 
exhibits a nonphysical oscillation
in the
low energy region. At somewhat higher numbers
of the retained oscillator functions this feature disappears
and  the phase shift becomes close to the exact one at any energy also with this  $R_{cut}$ value.

\begin{figure}[ht]
\centerline{\includegraphics*[scale=0.5]{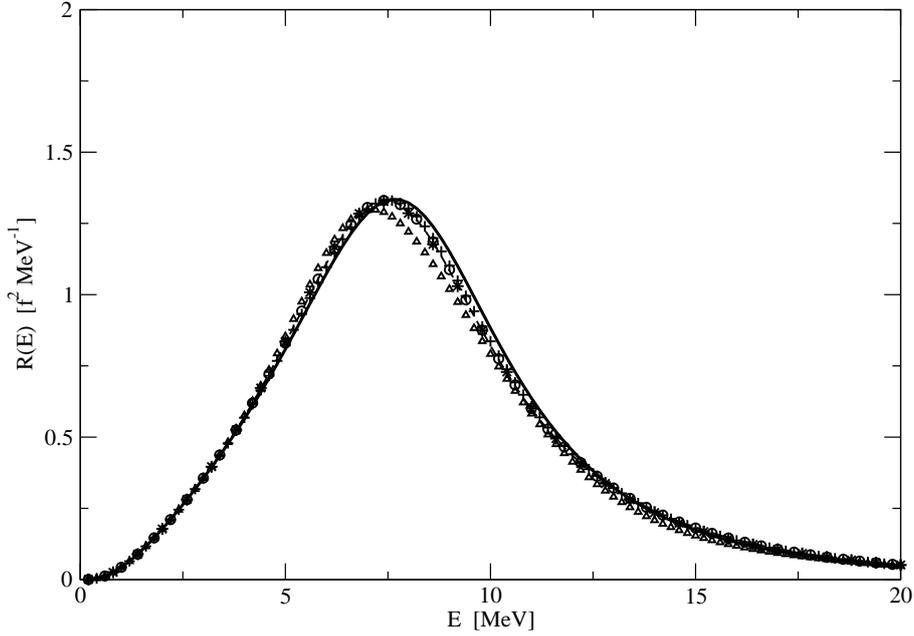}}
\caption{Comparison of the exact response function (full line) with those calculated with use of the oscillator basis.
Notation is as in Fig. 1.}   
\end{figure}

In Fig.~2 the dipole photodisintegration response function defined as 
\be R(E)=\left[\int_0^\infty r^2dr\,{\bar \psi}_E(r)r\psi_0(r)\right]^2\la{resp}\ee
is plotted. Here ${\bar \psi}_E=\psi_E\cdot(mk/\hbar^2)^{1/2}\cos\delta$ where $\psi_E$ is given by
Eq.~(\re{csp}), and $\psi_0$ is the  wave
function of the initial bound state 
normalized to unity and represented by 30 oscillator functions. 
The same comments as above apply to this case. Thus it is sufficient to retain seven basis
oscillator functions to represent the final continuum state.

The above results suggest that the maximum number of many--body 
oscillator quanta required in a many--body calculation to reach
convergence may be moderate.

The Coulomb interaction was disregarded in the above calculations but its inclusion could not
change the picture.  ME
of the Coulomb potential itself would not enter a calculation
due to the presence of the Coulomb wave functions in the corresponding ansatz of Eq.~(\re{mex})
or (\re{csp}) type. 
The Coulomb wave functions  in the internal region only are involved 
and the difference between their behavior in this region and the behavior of the spherical
Bessel functions  is of no importance.

\section{Shell--model wave functions of fragments excited with respect to center of mass}
The wave functions of fragments entering Eq.~(\re{sf}) are supposed to be taken from ncsm calculations. 
In fact, such calculations  give products
of  these wave functions and the lowest oscillator functions of the fragment centers of mass.
These products are provided in the form of expansions over the Slater determinants.
Let us denote such products as $X^{000}_{I_1M_1}$ and $X^{000}_{I_2M_2}$,
\ber X^{000}_{I_1M_1}(1,\ldots,A_1)=\phi^{I_1M_1}_1(1,\ldots,A_1)\Psi_{000}({\bf{\bar r}}_{cm}^{(1)}),\nonumber\\
X^{000}_{I_2M_2}(A_1+1,\ldots,A_1+A_2)
=\phi^{I_2M_2}_2(A_1+1,\ldots,A_1+A_2)\Psi_{000}({\bf{\bar r}}_{cm}^{(2)})\la{ggs}\eer
where  as in Eq.~(\re{sf})
$\phi^{I_1M_1}_1$ and $\phi^{I_2M_2}_2$
represent the eigenfunctions of  internal Hamiltonians having the angular momenta~$I_{1,2}$ and
their projections~$M_{1,2}$.    The  notation 
${\bf{\bar r}}_{cm}^{(1),(2)}$ stands for ${\bf r}_{cm}^{(1),(2)}\sqrt{A_{1,2}}/r_0$ where ${\bf r}_{cm}^{(1),(2)}$ 
are the center--of--mass vectors 
of the fragments. 

We shall need also products of the internal wave function of a fragment and the wave
function  of its center of mass in a given excited state.\footnote{After the 
present manuscript
was submitted I learned about Ref \cite{clust} where the states of Eq.  (\re{es}) form were constructed 
numerically using the center--of--mass 
creation and annihilation operators. In the present paper this is done analytically proceeding 
from Eq. (\re{up}) below. The oscillator cluster model (with no explicit antisymmetrization) 
in the frame of which these states are used in Ref. \cite{clust}
is quite different from the present scheme to calculate reactions.} 
Let us denote these  products as $X^{nlm}_{I_1M_1}$,
\be X^{nlm}_{I_1M_1}(1,\ldots,A_1)=\phi^{I_1M_1}_1(1,\ldots,A_1)\Psi_{nlm}({\bf{\bar r}}_{cm}^{(1)}),\la{es}\ee
($A_1>1$). We shall obtain them in the form of 
linear combinations of the Slater determinants as well. 
Use of these products in this form is a key element of the present approach. 

They are to be obtained in some ${\cal N}$ range,  $2n+l\le {\cal N}$, of numbers of the center of mass
oscillator quanta. If  
the states~(\re{ggs}) include components with the numbers of total many--body oscillator quanta 
up to some $N_{tot}$ then the states ~(\re{es}) will include the Slater determinants  with total many--body oscillator quanta  
up to $N_{tot}+{\cal N}$. For heavier fragments and at higher $2n+l$ values 
the amount of these determinants may be too large. 
However, the states~(\re{es}) with large $2n+l$ will  play only
a role of small corrections
in what follows.
Therefore, in the expansion of the states~(\re{es}) 
over the Slater determinants it may be acceptable to retain only the determinants  with total many--body oscillator quanta  
up to some $N_{tot}+{\cal N}_0$ value with ${\cal N}_0<{\cal N}$. 
Then at  $2n+l\le {\cal N}_0$ the procedure is exact while at $2n+l> {\cal N}_0$
 the components
of the states~(\re{ggs}) only with total many--body oscillator quanta 
up to $N_{tot}'$, such that
$N_{tot}'+2n+l=N_{tot}+{\cal N}_0$, will contribute to the result. 

This truncation does not violate the separation of the internal and center--of--mass motions. Indeed,  from the 
fact of the separation of these motions in Eqs.~(\re{ggs}) it follows
that the same separation, of course, takes place for separate components of the states (\re{ggs})
having given numbers of total many--body oscillator quanta. The procedure 
then merely either allows or forbids contributions of
these components to a resulting center--of--mass excited state. 
(Obviously, ${\cal N}_0$ 
 should be such that at least the minimal configuration of a state like (\re{ggs}) is not forbidden. This 
reads as $N_{tot}+{\cal N}_0\ge N_{min}+{\cal N}$ where $N_{min}$ is the number of
many--body oscillator quanta in the minimal configuration.)

Let us denote the internal Hamiltonian of a fragment
as $h$, the oscillator Hamiltonian
of its center of mass as $h_{cm}$, and the operators of the 
center--of--mass orbital momentum and its projection as $l^2_{cm}$ and $(l_z)_{cm}$.
The function $X^{nlm}_{I_1M_1}$ may be obtained as an eigenfunction of the operator 
$\lambda_1 h+\lambda_2 h_{cm}+\lambda_3l^2_{cm}+\lambda_4(l_z)_{cm}$ where 
$\lambda_i$ constants are not equal to zero and arbitrary otherwise. At the diagonalization of this operator
in the subspace of the Slater determinants with the number of quanta up to $N_{tot}+{\cal N}_0$ 
the eigenvalue pertaining to this eigenfunction is
\mbox{$\lambda_1 \epsilon_0+\lambda_2 (2n+l+3/2)+\lambda_3l(l+1)+\lambda_4m$} where  $\epsilon_0$ is
an approximate bound--state energy of the fragment. 

It seems simpler to compute the $X^{nlm}_{I_1M_1}$~states in another way. 
Let us write  $\Psi_{nlm}$ in terms of 
$\Psi_{000}$, see, e.g., \cite{mosh},
\be \Psi_{nlm}({\bf{\bar r}}_{cm}^{(1)})=\alpha_{nl}\sqrt{4\pi}
({\bs \eta}\cdot{\bs \eta})^n{\cal Y}_{lm}({\bs\eta})\Psi_{000}({\bf{\bar r}}_{cm}^{(1)})\la{up}\ee
where $\alpha_{nl}=(-1)^n[(2n+2l+1)!!(2n)!!]^{-1/2}$,
${\bs \eta}$ are the oscillator creation operators, 
\be {\bs \eta}=\frac{1}{\sqrt{2}}\left({\bf{\bar r}}_{cm}^{(1)}-\frac{d}{d{\bf{\bar r}}_{cm}^{(1)}}\right),\ee
and ${\cal Y}_{lm}$ are the solid harmonics. 
The $X^{nlm}_{I_1M_1}$~states are expressed in terms
of the $X^{000}_{I_1M_1}$~states in the same way as in Eq. (\re{up}). 
Then using Eq.~(\re{up}) one may obtain $X^{nlm}_{I_1M_1}$ from 
$X^{000}_{I_1M_1}$
via recurrence relations. First, one may employ the relation following from Eq.~(\re{up})
\be  X^{0l+1,l+1}_{I_1M_1}=(l+1)^{-1/2}\eta_{+}X^{0ll}_{I_1M_1}.\la{11}\ee
Here and below spherical components of vectors like $\eta_\pm=\mp 2^{-1/2}(\eta_x\pm i\eta_y)$ and $\eta_0=\eta_z$ 
are used.
Once $X^{0ll}_{I_1M_1}$ is constructed in the form of a linear combination of the 
Slater determinants, Eq.~(\re{11}) makes possible
to construct  in this form also $X^{0l+1,l+1}_{I_1M_1}$   using the relation
\be {\bs \eta}=\frac{1}{\sqrt{A_1}}\sum_{i=1}^{A_1}{\bs \eta}_i\la{expr}\ee
where ${\bs \eta}_i$ are the oscillator creation operators for separate nucleons,
\be {\bs \eta}_i=\frac{1}{\sqrt{2}}\left({\bf{\bar r}}_i-\frac{d}{d{\bf{\bar r}}_i}\right),\qquad {\bf{\bar r}}_i=
{\bf r}_i/r_0.\ee

In what follows only the states~(\re{es}) with $l$ values of the same parity will be required. In this 
connection,
the relation  similar to Eq.~(\re{11})
\be  X^{0l+2,l+2}_{I_1M_1}=[(l+1)(l+2)]^{-1/2}\eta_{+}^2X^{0ll}_{I_1M_1}\la{2}\ee
may also be useful in conjunction
with Eq.~(\re{expr}). One-- and two--body matrix
elements are then to be calculated.

Next, one obtains the $X^{nll}_{I_1M_1}$~states with $n\ne0$ as a combination of the Slater determinants 
applying the relation following from Eq.~(\re{up})
\be X^{n+1,ll}_{I_1M_1}=-[(2n+2l+3)(2n+2)]^{-1/2}{\bs \eta}\cdot{\bs \eta}X^{nll}_{I_1M_1}.\la{27}\ee
One needs to calculate one-- and two--body matrix
elements also in this case.

If the above discussed approximation is adopted and
the  $N_{tot}+{\cal N}_0$ maximal number of the total many--body oscillator quanta is reached then  
at performing each further step of
the procedure the Slater determinants 
that have this maximal number of quanta
are to be dropped in the right--hand sides of the corresponding
recursion relations, Eq.~(\re{27}) or (\re{11}), or (\re{2}).

In a certain case below  the  $X^{nlm}_{I_1M_1}$~states with $m<l$ are also required.
They can be constructed with the help of
the lowering operators.
Writing 
${\bf l}=-i{\bs \eta}\times{\bs \eta}^+$ where 
\[{\bs \eta}^+=\frac{1}{\sqrt{2}}\left({\bf{\bar r}}_{cm}^{(1)}+\frac{d}{d{\bf{\bar r}}_{cm}^{(1)}}\right)\]
one gets
\be X^{nl,m-1}_{I_1M_1}=2^{1/2}[(l+m)(l-m+1)]^{-1/2}(\eta_{-}\eta^+_0-\eta_0\eta^+_{-})
X^{nlm}_{I_1M_1}.\la{mi}\ee
Along with Eq.~(\re{expr}) one uses
the relations
\be {\bs \eta}^+=\frac{1}{\sqrt{A_1}}\sum_{i=1}^{A_1}{\bs \eta}_i^+,\quad {\bs \eta}_i^+=
\frac{1}{\sqrt{2}}\left({\bf{\bar r}}_i+\frac{d}{d{\bf{\bar r}}_i}\right)\ee
to calculate these states in the required form.
One--body contributions to the operator from Eq.~(\re{mi}) are the single--particle orbital 
momenta~$l_{-}^{(i)}$.

\section{relation between cluster and shell--model ME}

In this section we address the ME~(\re{mey}) for the case when the $\chi_k$~functions discussed in Sec. 2 
possess given $J$ values and correspond to the center of mass in the lowest
oscillator state. To signify these properties, these functions will be denoted as $\chi_k^{JM,0}$ in such a case.
The corresponding ME~(\re{mey}) do not depend on~$M$. We want to express them  in terms of the ME 
\be \left(\chi_k^{JM,0},{\hat O}X^{nlm}_{I_1M_1}(1,\ldots,A_1)X^{000}_{I_2M_2}(A_{1}+1,\ldots,A_{1}+A_{2})\right)
\equiv {\cal M}(I_1M_1I_2M_2,nlm,J)\la{me}\ee
where $M_1+M_2+m=M$. The states~$X^{nlm}_{I_1M_1}$  and $X^{000}_{I_2M_2}$ 
are of the form  of Eqs.~(\re{ggs}) and~(\re{es}) of the preceding section. 
The $X^{000}_{I_2M_2}$~state is directly provided by a bound--state 
ncsm calculation. At $A_1>1$ the  $X^{nlm}_{I_1M_1}$~states
are obtained from a ncsm bound state by means of the center of mass excitation as described in the preceding 
section.
 Both $X^{nlm}_{I_1M_1}$ and $X^{000}_{I_2M_2}$ are sums of the Slater determinants
as well as $\chi_k^{JM,0}$.

At $A_1=1$, i.e., in the case when the first fragment is a nucleon, the
function~$\phi^{I_1M_1}_1$ in Eqs.~(\re{ggs}) and (\re{es}) is to be replaced with the corresponding spin--isospin function 
of the nucleon. The $X^{nlm}_{I_1M_1}$ wave function is then merely the product of this 
spin--isospin function ($I_1=1/2$)  
and  $\Psi_{nlm}({\bs{\bar r}}_{cm}^{(1)})$.

Consider the quantities 
\be \sum_{M_1+M_2=M-m}C_{I_1M_1I_2M_2}^{S,M-m}{\cal M}(I_1M_1I_2M_2,nlm,J).\ee
Since  in the ME~(\re{me}) $\chi_k^{JM,0}$ is a state with a given total momentum and its projection
it is clear that these 
quantities are proportional to the corresponding 
ME in which $X^{nlm}_{I_1M_1}$ and $X^{000}_{I_2M_2}$ are coupled
to the total momentum according to the $((I_1I_2)Sl)J$~scheme, and the proportionality coefficient 
is the Clebsh--Gordan coefficient~$C^{JM}_{S,M-m,lm}$. 
Furthermore,  in Eq.~(\re{me}) the wave function pertaining to $X^{nlm}_{I_1M_1}$ includes  
$\Psi_{nlm}({\bf{\bar r}}_{cm}^{(1)})$ as a factor and that pertaining to $X^{000}_{I_2M_2}$ includes  
$\Psi_{000}({\bf{\bar r}}_{cm}^{(2)})$ as a factor.   One may write
\be \Psi_{nlm}({\bf{\bar r}}_{cm}^{(1)})\Psi_{000}({\bf{\bar r}}_{cm}^{(2)})=\sum_{n'l'NL}
\langle n'l'NL|nl00\rangle_l^{\varphi}\left[\Psi_{n'l'}({\bs{\bar \rho}})
\Psi_{NL}({\bf{\bar R}}_{cm})\right]_{lm}\la{trr}\ee
where $\langle n'lNL|nl00\rangle_l^{\varphi}$ are the oscillator brackets corresponding to the 
orthogonal transformation
\be {\bf{\bar r}}_{cm}^{(1)}={\bs{\bar \rho}}\cos\varphi+{\bf{\bar R}}_{cm}\sin\varphi,\qquad
{\bf{\bar r}}_{cm}^{(2)}=-{\bs{\bar \rho}}\sin\varphi+{\bf{\bar R}}_{cm}\cos\varphi\la{tr1}\ee
with $\cos\varphi=(A_2/A)^{1/2}$ and $\sin\varphi=(A_1/A)^{1/2}$.
Since in the ME~(\re{me}) 
the $\chi_k^{JM,0}$~wave function is proportional to $\Psi_{000}({\bf{\bar R}}_{cm})$, only the term 
from the sum in the right--hand side with $N=L=0$
and hence with $l'=l$ and $n'=n$  contributes to the result. 
Therefore, one has the relation
\ber\!\!\!\!\!\! \sum_{M_1+M_2=M-m}\!\!\!C_{I_1M_1I_2M_2}^{S,M-m}{\cal M}(I_1M_1I_2M_2,nlm,J)
=C^{JM}_{S,M-m,lm}\langle nl00|nl00\rangle_l^{\varphi}\left(\chi_k^{JM,0},{\hat O}
Z_{I_1I_2SJM}^{nl}\right)\!\!.\la{bf}\eer
The $m$  and $M-m$ values here are arbitrary.
This relation expresses the ME (\re{mey}) that contain cluster type wave functions in terms of 
the ME (\re{me}) that involve only oscillator orbitals. Similar relations have been derived, e.g.,  in 
\mbox{Refs.~\cite{Nav,Smi,Rot,Smi1}}, their  differences with Eq.~(\re{bf}) refer to dealing with angular momenta.
A more general relation is given in Appendix~B.

(One has 
$\langle nl00|nl00\rangle_l=(A_2/A)^{(2n+l)/2}$ in Eq.~(\re{bf}).
This known relation follows, e.g., from Eq.~(\re{br0}) below.)

One may choose $m=l$ in Eq.~(\re{bf}). With
 this choice, the Clebsh--Gordan coefficient in the right--hand side of Eq. (\re{bf}) 
is different from zero at least
when $M=J$ is chosen. Indeed,~\cite{var}
\be C^{JJ}_{S,J-l,ll}=(-1)^{l+S-J}\left[\frac{(2l)!(2J+1)!}{(l+S+J+1)!(l-S+J)!}\right]^{1/2}.\ee
Thus the ME sought for may be computed, e.g., from the relation
\ber\!\!\!\! \left(\chi_k^{JM,0},{\hat O}Z_{I_1I_2SJM}^{nl}\right)=\left[C^{JJ}_{S,J-l,ll}\,
\langle nl00|nl00\rangle_l^{\varphi}\right]^{-1}
\!\!\!\!\!\!\sum_{M_1+M_2=J-l}C_{I_1M_1I_2M_2}^{S,J-l}{\cal M}(I_1M_1I_2M_2,nll,J).\la{bf1}\eer
It is clear that in the above relations it is expedient to choose the lighter of the two fragments 
to be the fragment number one in the notation we use.  
A similar type relation can be written also in the case when the $\chi_k$~states do not possess given momentum~$J$ but 
still correspond to the center of mass of the whole system in the ground state.

\section{Matrix elements involving channel functions}

The coefficients of the dynamic equations of Sec. 2 are the bound--bound ME and the bound--free ME, 
i.e., those which include 
the localized ncsm functions  and 
non--localized channel functions of Eqs.~(\re{ch}) and (\re{ch1}). Thus, finally we need to calculate the bound--free ME
using the considerations above. 
These ME are  of the form
\be\left(\chi_k,[H-E]{\cal A}_i\varphi_if_i(\rho_i)\Psi_{000}({\bf{\bar R}}_{cm})\right)\la{str}\ee
where ${\cal A}_i$ is the antisymmetrizer (\re{a}), $\varphi_i$ is the surface function (\re{sf}), 
$f_i$ one of the functions~$f^{(0),(1)}_i$ from Eq.~(\re{z1})
or functions~$f^{\pm}_i$ from Eq.~(\re{f}), and
the notation
$\chi_k$ is as above. Using the fact that $\chi_k$ 
are antisymmetric with respect to permutations one may rewrite 
the expression (\re{str}) as
\ber\bigl({\cal A}_i^\dag\chi_k,[H-E]\varphi_if_i(\rho_i)\Psi_{000}({\bf{\bar R}}_{cm})\bigr)
=\nu_i^{1/2}\left(\chi_k,[H-E]\varphi_if_i(\rho_i)\Psi_{000}({\bf{\bar R}}_{cm})\right).\la{str1}\eer
Thus the task of antisymmetrization of the channel functions is removed. 

The ME  in the right--hand side of Eq.~(\re{str1}) is of the structure
\ber \left(\chi_k,[H-E]
\Bigl[\left[\phi_1^{I_1}(1,\ldots,A_{1})\phi_2^{I_2}(A_{1}+1,\ldots,A_{1}+A_{2})\right]_{S}
Y_{l}({\hat{\bs \rho}})\Bigr]_{JM}
 f(\rho)\Psi_{000}({\bf{\bar R}}_{cm})\right).
\la{38}\eer
The channel subscript~$i$ is omitted here and below. An efficient way to calculate it
 is as follows. One uses the $H-E$~operator 
in the form~(\re{dec}) and one treats the fragment wave functions as  being exact. Then
the expression~(\re{38}) becomes 
\ber \left(\chi_k,\Bigl[\left[\phi_1^{I_1}\phi_2^{I_2}\right]_{S}
Y_{l}({\hat{\bs \rho}})\Bigr]_{JM}{\tilde f(\rho)}\Psi_{000}({\bf{\bar R}}_{cm})\right)\nonumber\\
+\left(\chi_k^{JM,0},[V_{ext}^{nucl}+V_{ext}^{coul}-{\bar V}_{ext}^{coul}(\rho)]
\Bigl[\left[\phi_1^{I_1}\phi_2^{I_2}\right]_{S}
Y_{l}({\hat{\bs \rho}})\Bigr]_{JM}f(\rho)\Psi_{000}({\bf{\bar R}}_{cm})\right)\la{sme}\eer
where
\be {\tilde f(\rho)}=\left[-\frac{\hbar^2}{2\mu}\left(\frac{d^2}{d\rho^2}+\frac{2}{\rho}\frac{d}{d\rho}-
\frac{l(l+1)}{\rho^2}\right)+{\bar V}_{ext}^{coul}(\rho)-E_{rel}\right]f(\rho).\ee
The function~${\tilde f(\rho)}$ is localized. Let us approximate it
by its  truncated expansion over the oscillator functions. Then the first of the ME in Eq.~(\re{sme}) turns to a
sum of the ME 
\be \left(\chi_k,Z_{I_1I_2SJM}^{nl}\right)\la{mey1}\ee
of the type of Eq.~(\re{mey}).
If $f=f^{(1)}$  one has ${\tilde f}=0$ and this contribution is absent. 

In the second of the ME in Eq.~(\re{sme}) only values of $f(\rho)$ in a limited $\rho$~range 
contribute to the result. Therefore, for computation of this ME one may use an approximation of $Y_{lm}f$ 
by a sum of oscillator 
functions. 
For this purpose, $f(\rho)$ is approximated with a linear combination $\sum_{n=0}^{n_{max}}c_nR_{nl}$
of the radial parts $R_{nl}(\rho)$ of the 
oscillator functions $\Psi_{nlm}({\bs {\bar \rho}})$ used above.   This can be done via minimization of the quantity
\be\int_0^\infty d\rho\,\omega(\rho)\Bigl|f(\rho)-\sum_{n=0}^{n_{max}}c_nR_{nl}(\rho)\Bigr|^2\la{ap}\ee 
with respect to the $c_n$~coeficients, $\omega(\rho)$ being a localized positive weight function. 
As a result, the contribution of the $V_{ext}^n+V_{ext}^c$ term to the ME turns to a    
sum of the ME 
\be \left(\chi_k,[V_{ext}^{nucl}+V_{ext}^{coul}]Z_{I_1I_2SJM}^{nl}\right)\la{mey2}\ee
again of the type of Eq.~(\re{mey}). To calculate 
the contribution of the ${\bar V}_{ext}^{coul}(\rho)$ term one may represent the arising product
\mbox{${\bar V}_{ext}^{coul}(\rho)\sum_{n=0}^{n_{max}}c_nR_{nl}(\rho)$} as a sum of the functions~$R_{nl}(\rho)$ 
minimizing the quantity similar to  (\re{ap}) at the same $\omega(\rho)$.
As a result, 
this contribution takes the above form of a sum of Eq.~(\re{mey1}) type ME. 
(Thus, besides the total number of basis functions, the parameters of a calculation with respect to which its stability 
is to be checked  are the $n_{max}$~type numbers
and possibly the above defined ${\cal N}_0$~numbers.)

In what follows, let us first consider the case when the $\chi_k$ states possess given $J$~values and correspond to 
the center of mass of the whole system in the ground state, i.e. $\chi_k^{JM,0}$~states in the above notation. 
Then
applying the relation (\re{bf}) or (\re{bf1}) one reduces the ME (\re{mey1}) and (\re{mey2}) to a sum of the quantities
of Eq.~(\re{me}) type with ${\hat O}=I$ or ${\hat O}=V_{ext}^{nucl}+V_{ext}^{coul}$. 
Each of these quantities  is the sum  of ME that contain the products of 
three Slater determinants
entering, respectively, the $\chi_k^{JM,0}$~basis function
and the $X^{nlm}_{I_1M_1}$ and $X^{000}_{I_2M_2}$~fragment wave functions. 

In the  mentioned ${\hat O}=I$ case
consider such an ME in which the first of the mentioned determinants is 
constructed of the oscillator orbitals~$\psi_l(i)$ with \mbox{$1\le i\le A_1+A_2$} and \mbox{$l=\{l_1,\ldots,l_{A_1+A_2}\}$}, 
the second determinant is 
constructed of the oscillator orbitals~$\psi_m(j)$ with \mbox{$1\le j\le A_1$ and $m=\{m_1,\ldots,m_{A_1}\}$}, 
and the third one is 
constructed of the oscillator orbitals~$\psi_n(k)$ with $A_1+1\le k\le A_1+A_2$ and $n=\{n_1,\ldots,n_{A_2}\}$.
The oscillator orbitals are assumed to be orthonormalized. 
The ME is calculated  performing the Laplace
expansion of the first of the mentioned determinants  over the minors pertaining to the $\psi_l(i)$~orbitals 
with $1\le i\le A_1$.
The ME is different from zero only 
if the $\{l\}$ set of orbitals coincides with the $\{\{m\},\{n\}\}$ set. (This is only possible
if all the orbitals belonging to the $\{m\}$~set differ from those belonging to the $\{n\}$~set.) 
In this case it is equal to
$\pm A_1!A_2!$ and the sign is governed by the simple rule.

In the mentioned ${\hat O}=V_{ext}^{nucl}+V_{ext}^{coul}$~case suppose that $V_{ext}^{nucl}+V_{ext}^{coul}$
is the sum of two--nucleon interactions~$V(ij)$.
This sum may be replaced with $A_1A_2V(ij)$ where $V(ij)$ is the interaction between a nucleon belonging to
one of the fragments and  a nucleon belonging to the other fragment.  Then
applying the relation~(\re{bf}) or (\re{bf1}) the contribution~(\re{mey2}) is  
reduced to a sum of ME between a Slater determinant pertaining to $\chi_k^{JM,0}$ and a product
of Slater determinants pertaining  to $X^{nlm}_{I_1M_1}$  and
$X^{000}_{I_2M_2}$. These ME are calculated with the help of the same Laplace
expansion as above. They are of the form 
\ber \Bigl(\det\left[\psi_{l_1'}(1),\ldots,\psi_{l_{A_1}'}(A_1)\right]
\det\left[\psi_{l_{A_1+1}'}(A_1+1),\ldots,\psi_{l_{A_1+A_2}'}(A_1+A_2)\right],\nonumber\\
V(ij)\det\left[\psi_{m_1}(1),\ldots,\psi_{m_{A_1}}(A_1)\right]
\det\left[\psi_{n_{A_1+1}}(A_1+1),\ldots,\psi_{n_{A_1+A_2}}(A_1+A_2)\right]\Bigr)\la{52}\eer
where $1\le i\le A_1$ and $A_1+1\le j\le A_1+A_2$, and the 
sets  $\{l_1',\ldots,l_{A_1}'\}$ and $\{l_{A_1+1}',\ldots,l_{A_1+A_2}'\}$
are subsets of the $\{l_1,\ldots,l_{A_1+A_2}\}$ set.
These ME may not vanish only if the latter subsets differ, respectively, from the $\{m\}$~set and the $\{n\}$~set
by not more than one orbital. In such cases the set of the $\{\{m\},\{n\}\}$~orbitals 
differs from the $\{l\}$ set by not more than  two orbitals
and these two orbitals cannot belong to the same $\{m\}$ or $\{n\}$~set. 
(This is only possible
if in the $\{m\}$ ~set not more than two orbitals are the same as in the $\{n\}$~set.)
The ME~(\re{52}) is of the structure similar to
that of one--body operators.  
Therefore, it is reduced in the usual way (depending on whether the corresponding orbitals are the same or not)
to the standard two--body ME like
\[ \bigl(\psi_l(i)\psi_{l'}(j),V(ij)\psi_m(i)\psi_n(j)\bigr).\]
ME of three--nucleon interactions that contribute to $V_{ext}^{nucl}$ are calculated in a similar way. 

If $\chi_k$ is a combination of Slater determinants which does not possess definite quantum numbers of 
 the center of mass of the whole system then the simplification of the preceding section is not applicable anymore. But
then, just as above, the ME~(\re{str}) still can be written as the sum of quantities of Eq.~(\re{mey}) form 
where ${\hat O}$ is either a unit operator
or an operator of two-- or three--body inter--fragment interaction. While above these quantities were expressed
in terms of the contributions (\re{me}), now they
 may be represented as the sums 
of the contributions of the form
\be \left(\chi_k,{\hat O}X^{n_1l_1m_1}_{I_1M_1}(1,\ldots,A_1)
X^{n_2l_2m_2}_{I_2M_2}(A_{1}+1,\ldots,A_{1}+A_{2})\right)\la{mme}\ee  
where $X^{n_1l_1m_1}_{I_1M_1}$ and $X^{n_2l_2m_2}_{I_2M_2}$ are obtained  via the center--of--mass excitaion
applied to ncsm wave functions of the fragments as in Sec. 3. (In the $X^{n_2l_2m_2}_{I_2M_2}$ case one proceeds
in the same way as described in Sec. 3 as to $X^{n_1l_1m_1}_{I_1M_1}$.)
To this aim, let us directly transform the cluster $Z_{I_1I_2SJM}^{nl}$~state to the form of the 
sum of products of  $X^{n_1l_1m_1}_{I_1M_1}$ and 
$X^{n_2l_2m_2}_{I_2M_2}$.
This is achieved via the transformation 
\be \Psi_{nlm}({\bs{\bar \rho}})\Psi_{000}({\bf{\bar R}}_{cm})=\sum_{n_1l_1n_2l_2}
\langle n_1l_1n_2l_2|nl00\rangle_l^{\varphi'}
\left[\Psi_{n_1l_1}({\bf{\bar r}}_{cm}^{(1)})
\Psi_{n_2l_2}({\bf{\bar r}}_{cm}^{(2)})\right]_{lm}\la{trr1}\ee
which corresponds to the coordinate transformation reverse to Eq.~(\re{tr1}) so that $\varphi'$ equals~$-\varphi$
from there. 
(The brackets~$\langle n_1l_1n_2l_2|nl00\rangle_l^{\varphi'}$ are as follows 
\ber \langle n_1l_1n_2l_2|nl00\rangle^{\varphi'}_l=\cos^{2n_1+l_1}\varphi'\sin^{2n_2+l_2}\varphi'
\nonumber\\
\times(-1)^l
\left[(2l_1+1)(2l_2+1)\right]^{1/2}\left(\begin{array}{ccc}l_1&l_2&l\\0&0&0\end{array}\right)
\frac{\alpha_{n_1l_1}\alpha_{n_2l_2}}{\alpha_{nl}}\la{br0}\eer
where the $\alpha$~coefficients are defined in Eq.~(\re{up}).
This relation is obtained in Appendix C.)

It is clear that the quantities~(\re{mme}) are calculated in the same
way as the quantities~(\re{me}) above. In the case we consider now,
computations are more lengthy. 

In conclusion,  reaction observables can be computed  
from simple linear equations with use of the Slater determinants. The equations do not
include free--free matrix elements. 
Antisymmetrization between nucleons belonging to different fragments is not required.
In model examples, the equations lead to rather precise results.   
The required bound--free ME 
are calculated in a universal way.   
Their computation somewhat resembles that of the  ncsm bound--state ME.
 One may expect that the 
convergence rates of reaction observables in the present scheme
should in general be at the level of the convergence rates
of typical bound--state observables in ncsm. 
 
A method applicable 
for reactions at higher energy, when channels with three or more fragments are open,
may be developed relying on the considerations of the present work in conjunction with the
integral transform approach, see \cite{28,rev,29,30}.

The main objects of that approach are response--like functions $R(E)$ and their integral transforms $\Phi(\sigma)$.
The transforms  $\Phi(\sigma)$ are obtained via solving a many--body problem that does not require the specification of 
reaction channels and resembles a bound--state calculation. The response--like functions $R(E)$  
are subsequently obtained with the help of inversion of the transforms, that is via solving integral equations of the type
\be \Phi(\sigma)=\int_{E_{thr}}^\infty dE\,K(\sigma,E)R(E).\la{inteq}\ee
The Lorentz kernel \mbox{$K(\sigma_R+i\sigma_I,E)=[(\sigma_R-E)^2+\sigma_I^2]^{-1}$}
 is a good choice \cite{31} in many cases.

The accuracy with which $R(E)$ can be found from Eq. (\re{inteq}) is governed by the accuracy of the input $\Phi(\sigma)$.
Finding $R(E)$  with a reasonable accuracy at low energy $E$  requires the most accurate $\Phi(\sigma)$. This is because
of possible narrow resonances at low energy and fast variations of $R(E)$ in the vicinity of the lowest threshold $E_{thr}$.
It may be profitable to introduce the energy $E_0$ below which $R(E)$ is calculated directly employing continuum wave 
functions. The energy $E_0$ lies below the three--cluster breakup threshold and the method of the present work may therefore
 be applied to compute these wave functions. And at energies $E>E_0$ one calculates $R(E)$ from the modified
Eq.  (\re{inteq}),
\[ {\Phi}'(\sigma)=\int_{E_0}^\infty dE\,K(\sigma,E)R(E),\qquad
{\Phi}'(\sigma)=\Phi(\sigma)-\int_{E_{thr}}^{E_0} dE\,K(\sigma,E)R(E).\]

Support from RFBR Grant No. 18--02--00778 is acknowledged.
 
\appendix
\section{Matrix elements  involving the Laguerre--type basis}
Below the ME of the $H-E$~operator between the functions
entering Eq.~(\re{mex}) are listed. Subscripts like $m$ refer to the 
functions (\re{basl}).
The ME of the radial kinetic energy between the functions~(\re{basl}) for a state with an orbital momentum~$l$ are 
as follows,
\ber T_{mm'}=\frac{\hbar^2}{2\mu d^2}\left[\frac{(m_<+1)(m_<+2)}{(m_>+1)(m_>+2)}\right]^{1/2}
\left[\frac{1}{2}-\frac{\delta_{mm'}}{4}+\frac{m_<}{3}+\frac{l(l+1)(3m_>-m_<+3)}{6}\right]\eer
where $m_>=\max(m-1,m'-1)$ and $m_<=\min(m-1,m'-1)$. 
The ME of the exponential potential between the functions~(\re{basl}) are calculated as finite sums with
the help of the known relation~\cite{grad}, Eq.~7.414 (4).  

To calculate the ME of the $H-E$~operator between the functions~$\chi_m$ (\re{basl}) and 
the two scattering functions entering Eq.~(\re{mex}) the following scheme is convenient. All these ME 
are readily obtained from the integrals of the form
\be \int_0^\infty dt e^{-zt}tL_{m-1}^2(t)\la{i1}\ee
with $z=a+ib$, $a>0$, and $b=\mp ikd$. These integrals  can be calculated with the help of the following simple relation 
\be \int_0^\infty dt e^{-zt}L_n^1(t)=1-\left(\frac{z-1}{z}\right)^{n+1}.\la{i2}\ee
We derived it from an expression~\cite{grad}, Eq.~7.414 (5), involving the general Laguerre polynomial~$L_n^\alpha(t)$. 
The integrals~(\re{i1}) are obtained from Eq.~(\re{i2}) via the recurrence
relation \mbox{$tL_{m-1}^2(t)=mL_{m-1}^1(t)-(m+1)L_{m}^1(t)$}.

\section{More general relation between cluster  and shell--model matrix elements }
In the notation of Sec. 3 let us define the ME
\ber {\cal M}(I_1M_1I_2M_2,n_1l_1m_1,n_2l_2m_2,J)
\nonumber\\
\equiv
\left(\chi_{JM},{\hat O}X^{n_1l_1m_1}_{I_1M_1}(1,\ldots,A_1)X^{n_2l_2m_2}_{I_2M_2}(A_{1}+1,\ldots,A_{1}+A_{2})\right).
 \la{me1}\eer
The ME (\re{mey}) sought for may be computed in terms of these ME as follows, 
\ber \left(\chi_{JM},{\hat O}Z_{I_1I_2SJM}^{nl}\right)=
\left[C_{SM_Slm}^{JM}\langle nl00|n_1l_1n_2l_2\rangle^\varphi_l\right]^{-1}\nonumber\\ \times\sum_{M_1M_2m_1m_2}
C_{I_1M_1I_2M_2}^{SM_S}C_{l_1m_1l_2m_2}^{lm}{\cal M}(I_1M_1I_2M_2,n_1l_1m_1,n_2l_2m_2,J).\la{gr}\eer
Here $M_S+m=M$, $2n_1+l_1+2n_2+l_2=2n+l$, and $M_S,m,n_1,l_1,n_2$, and $l_2$ are arbitrary otherwise.
The oscillator brackets entering here are defined in accordance with the transformation
\be\left[\Psi_{n_1l_1}({\bf{\bar r}}_{cm}^{(1)})\Psi_{n_2l_2}({\bf{\bar r}}_{cm}^{(2)})\right]_{lm}=\sum_{n'l'NL}
\langle n'l'NL|n_1l_1n_2l_2\rangle_l^{\varphi}\left[\Psi_{n'l'}({\bs{\bar \rho}})
\Psi_{NL}({\bf{\bar R}}_{cm})\right]_{lm}\la{brr}\ee
and the relations (\re{tr1}) are implied.
 Eq. (\re{gr}) is obtained similarly
to Eq. (\re{bf}) and at 
$n_2=l_2=0$ it turns to it. The bracket
$\langle nl00|n_1l_1n_2l_2\rangle^\varphi_l$ is given by Eqs. (\re{br0}) and (\re{symm}). 
Eq. (\re{gr}) is more involved than Eq. (\re{bf})  but (e.g. in the case when both fragments
are the $\alpha$--particles) it may be employed 
 to reduce the maximum center--of--mass excitations of the fragments required at a given $2n+l$ value.

\section{The oscillator bracket $\langle n_1l_1n_2l_2|nl00\rangle^{\varphi}_l$.}

The definition of the oscillator brackets adopted in the paper
is as follows. Suppose that, as in Eq.~(\re{tr1}), 
\be {\bf x}={\bf x}'\cos\varphi+{\bf y}'\sin\varphi,\qquad{\bf y}=-{\bf x}'\sin\varphi+{\bf y}'\cos\varphi.\la{tt}\ee
Then one has
\be \left[\Psi_{n_1l_1}({\bf x})\Psi_{n_2l_2}({\bf y})\right]_{lm}=\sum_{n_1'l_1'n_2'l_2'}
\langle n_1'l_1'n_2'l_2'|n_1l_1n_2l_2\rangle^{\varphi}_{l}\left[\Psi_{n_1'l_1'}({\bf x}')\Psi_{n_2'l_2'}({\bf y}')\right]_{lm}.
\la{gbr}\ee 

To get the $\langle n_1l_1n_2l_2|nl00\rangle^{\varphi}_l$ bracket we shall 
use the
symmetry relation \be\langle n_1l_1n_2l_2|nl00\rangle^{\varphi}_l=
\langle nl00|n_1l_1n_2l_2\rangle^{-\varphi}_l\la{symm}\ee and calculate the bracket in its right--hand side.

Eq.~(\re{gbr}) is a relation between the polynomials of six variables.   
It leads to relations between their
components  of a given power. 
Writing at $\varphi\rightarrow-\varphi$ the relation for  the highest
power polynomials one gets 
\ber \alpha_{n_1l_1}\alpha_{n_2l_2}x^{2n_1}y^{2n_2}
\left[{\cal Y}_{l_1}({\bf x}){\cal Y}_{l_2}({\bf y})\right]_{lm}\nonumber\\=
\sum_{n_1'l_1'n_2'l_2'}
\langle n_1'l_1'n_2'l_2'|n_1l_1n_2l_2\rangle^{-\varphi}_{l}\,\alpha_{n_1'l_1'}\alpha_{n_2'l_2'}
(x')^{2n_1'}(y')^{2n_2'}\left[{\cal Y}_{l_1'}({\bf x}'){\cal Y}_{l_2'}({\bf  y}')\right]_{lm}
\la{e2}\eer
where in Eqs.~(\re{tt}}) the replacement $\varphi\rightarrow-\varphi$ 
is implied.
Let us write ${\bf x}$ and ${\bf y}$ here in terms of ${\bf x}'$ and ${\bf y}'$ and then
take the  $y'\rightarrow0$ limit.  This gives
${\bf x}={\bf x}'\cos\varphi$ and  ${\bf y}={\bf x}'\sin\varphi$. In Eq.~(\re{e2})
only the term   with \mbox{$n_2'=l_2'=0$} and hence $l_1'=l$ and $n_1'=n$ survives in this limit, $n$ being defined via
the relation \mbox{$2n_1+l_1+2n_2+l_2=2n+l$}.
Comparing the left-- and right--hand sides of the arising equality 
(which is convenient to do at the ${\bf x}'$~vector directed along the $z$~axis) one comes to Eq.~(\re{br0}).


\begin{thebibliography}{00}
\bibitem{Bar} B.R. Barrett, P. Navr\'atil, and J.P. Vary, Progr. Part. Nucl. Phys. {\bf 69}, 131 (2013).
\bibitem{Nav} P. Navr\'atil, Phys. Rev. C {\bf 70}, 054324 (2004).
\bibitem{qn} S. Quaglioni and P. Navr\'atil, Phys. Rev. C {\bf 79}, 044606 (2009).
\bibitem{bar} S. Baroni, P. Navr\'atil, and S. Quaglioni, Phys. Rev. C {\bf 87}, 034326 (2013).
\bibitem{nq} P. Navr\'atil and S. Quaglioni, Phys. Rev. C {\bf 83}, 044609 (2011).
\bibitem{qnrh} S. Quaglioni, P. Navr\'atil, R. Roth, and W. Horiuchi, J. Phys.:
Conf. Ser. {\bf 402}, 012037 (2012).
\bibitem{pap}G. Papadimitriou, J. Rotureau, N. Michel, M. Ploszajczak, and B. R. Barrett, Phys. Rev. C {\bf 88}, 044318 (2013).
\bibitem{s} A.M. Shirokov, A.I. Mazur, I.A. Mazur, and J.P. Vary, Phys. Rev. C {\bf 94}, 064320 (2016)
and references therein;  A. M. Shirokov, A. I. Mazur, I. A. Mazur, E. A. Mazur, I. J. Shin, Y. Kim, L. D. Blokhintsev, 
and J. P. Vary, Phys. Rev. C {\bf 98}, 044624 (2018).
\bibitem{ze} M.V. Zhukov and V.D. \'Efros, Yad.  Fiz.  {\bf 14}, 577
(1971)  [Sov. J. Nucl. Phys. {\bf 14}, 322 (1972)].
\bibitem{lad} K. Lad\'anyi and T. Szondy, Nuovo Cim. B {\bf 5}, 70 (1971).
\bibitem{saad} Y. Saad, {\it Iterative Methods for Sparse Linear Systems} (SIAM, Philadelphia PA, 2003).  
\bibitem{shir} A.M. Shirokov, Phys. At. Nucl. {\bf 69}, 1030 (2006).
\bibitem{gl} D. H. Gl\"okner and D. R. Lawson, Phys. Lett. B {\bf 53}, 313 (1974).
\bibitem{gw} M. Goldberger, K. Watson, {\it Collision Theory} (Dover Publications, N.Y., 2004).
\bibitem{del} L.M. Delves, Nucl. Phys. {\bf 29}, 326 (1962).
\bibitem{mo} H. Morawitz, J. Math. Phys. {\bf 11}, 649 (1970).
\bibitem{Ga} V. Galitzki, B. Karnakov, V. Kogan, and V. Galitzki, Jr, {\it Exploring Quantum Mechanics} (Oxford University
Press, 2013).
\bibitem{su} Y. Suzuki, W. Horiuchi, and D. Baye, Progr. Theor. Phys. {\bf 123}, 547 (2010).
\bibitem{es} V.D. Efros, W. Leidemann, and V.Yu. Shalamova, Few--Body Syst. (to be published).
\bibitem{clust}  K. Kravvaris and A. Volya, Phys. Rev. Lett. {\bf 119}, 062501 (2017).
\bibitem{mosh} M. Moshinsky, {\it The harmonic oscillator in modern physics; from atoms to quarks} (Gordon and Breach, 
N.Y., 1969). 
\bibitem{Smi} Yu.F. Smirnov and D. Chlebowska, Nucl. Phys. {\bf 26}, 306 (1961).
\bibitem{Rot} I. Rotter, Nucl. Phys. {\bf A122}, 567 (1968); {\it ibid.} {\bf A135}, 378 (1969).
\bibitem{Smi1} Yu.F. Smirnov and Yu.M. Tchuvil'sky, Phys. Rev. C {\bf 15}, 84 (1977).
\bibitem{var} D.A. Varshalovich, A.N. Moskalev, and V.K. Khersonskii, {\it Quantum Theory of Angular Momentum} 
(World Scientific, Singapore, 1988).
\bibitem{28} V.D. Efros, Yad.
Fiz. {\bf 41},  1498 (1985) [Sov. J. Nucl. Phys. {\bf 41}, 949 (1985)].
\bibitem{rev}V.D. Efros, W. Leidemann, 
G. Orlandini, and N. Barnea, J. Phys. G: Nucl. Part. Phys. {\bf 34},  R459 (2007); V.D. Efros, Phys. At. Nucl. 
{\bf 62},  1833 (1999) (arXiv: nucl-th/9903024).
\bibitem{29}P. Navr\'atil, S. Quaglioni, I. Stetcu, and B.R. Barret, J. 
Phys. G: Nucl. Part. Phys. {\bf 36}, 083101
(2009); M.D. Schuster, S. Quaglioni, C.W. Johnson, E.D. Jurgenson, and P. Navr\'atil, Phys. Rev. C {\bf 92}, 014320 (2015).
\bibitem{30}V.D. Efros, 
Phys. Rev. E {\bf 86}, 016704 (2012).
\bibitem{31}V.D. Efros, W. Leidemann, and G. Orlandini, 
Phys. Lett. B {\bf 338},  130 (1994).
\bibitem{grad}
I.S. Gradshteyn and I.M. Ryzhik, {\it Table of Integrals, Series, and Products} (Academic Press,  Amsterdam, 2007).

\end{thebibliography}
\end{document}